\documentclass{ws-procs9x6}            
\usepackage{relsize,color}

\begin{document}
\title{Luminosity constraint and entangled solar neutrino signals}
\author{Francesco Vissani}

\address{INFN, Laboratori Nazionali del Gran Sasso, 
Assergi, L'Aquila, Italy\\
$^*$E-mail: vissani@lngs.infn.it}

\begin{abstract}
The current frontier of solar neutrino physics is the observation of CNO neutrinos.   The signals of 
CNO and PP neutrinos are however entangled: 
In fact, the pp-flux can be known precisely only if CNO-flux is quantified; the interpretation of the gallium experiments 
depends upon both fluxes; a precise knowledge of pep-flux is a precondition to extract the CNO-neutrino signal with Borexino.
The luminosity constraint plays an increasingly important r\^ole: we present it in a new form, improving the expression obtained by J.~Bahcall.\cite{lum}

\end{abstract}


\bodymatter

\section{Introduction}\label{sec1}

The standard solar model (SSM) has always been
a crucial tool to understand the Sun.\cite{jb}  The expected solar neutrinos fluxes $\Phi_i$ are expressed as
\begin{equation} \label{ntn}
\Phi_i= \varphi_i \times  \frac{10^{\alpha_i}}{\mbox{cm$^2$s}} \mbox{ where }
{i=\mbox{\small pp,Be,pep,B,hep,N,O,F}}
\end{equation}
Here, the index $i$ runs over the 5 neutrino fluxes of the PP-chain and the 3 neutrino fluxes of the CNO-cycle,  
$\alpha_i$ are fixed exponents and $\varphi_i$ adimensional coefficients.
The SSM  predictions of the last 30 years,  
including the recent ones   
overviewed  by A.~Serenelli,\cite{dre} 
are given in \tref{tab1}.
Note that, \newline
1) the exponents $\alpha_i$ never changed, i.e., the orders of magnitude are stable;\\
2) the coefficients $\varphi_i$ of the CNO-cycle changed considerably. \\
The CNO-cycle 
is subdominant but also poorly known. In the most recent SSM, 
$\varphi_{\mbox{\tiny N}}$ goes from  3.20 (upper 1$\sigma$ range of GS98) to   1.75  (lower 1$\sigma$ range of AGSS09). The ensuing span $2.48\pm 0.73$ is compatible with 
$\varphi_{\mbox{\tiny N}}=0$ 
 at 3.4$\sigma$, i.e.,  CNO-neutrinos could be absent at 0.06\% CL. 
Thus,\\[0.4ex]
\centerline{\em a few $\sigma$ measurement will already impact on present knowledge.}

\begin{table}[t]
\tbl{Theoretical predictions for $\varphi_i$, i.e., for solar neutrino fluxes in 4 SSM. 
For non-symmetric BP2000 uncertainties, maximum errors are conservatively quoted.\label{tab1}}
{\begin{tabular}{ c || ccccc | ccc }
&\multicolumn{5}{c}{The PP-chain} &\multicolumn{3}{c}{The CNO-cycle}\\
\toprule
                         identification index  $i$   & {pp} & ${\mbox{Be}}$ &  {pep} & ${\mbox{B}}$  & hep &
                       ${\mbox{N}}$ & ${\mbox{O}}$ & ${\mbox{F}}$ \\  
                                                       exponent $\alpha_i$              & 10  & $9$ & $8$ & $6$  & $3$ & ${8}$ & ${8}$  & ${6}$  \\  
                        \toprule
                          BU1988 \cite{bu88}            & 6.0 & 4.7 &  1.4 & 5.8  & 7.6 & 6.1 & 5.2 & 5.2  \\ 
 $\mbox{\tiny errors}$              & 0.1  & 0.7 &  0.1 & 2.1  & 7.6 & 3.1 & 3.0  & 2.4 \\   \hline
BP2000  \cite{bp00}           & 5.95& 4.77 &  1.40 & 5.05  & 9.3 & 5.48 & 4.80 & 5.63  \\ 
 $\mbox{\tiny errors}$              & 0.06  & 0.48 &  0.02 & 1.01  & 9.3 & 1.15 & 1.20  & 1.41 \\   \hline
B16-GS98 \cite{vin}              &  5.98 & 4.93 &  1.44 & 5.46  & 7.98 & 2.78 & 2.05  & 5.29 \\ 
      $\mbox{\tiny errors}$           & 0.04 & 0.30 &  0.01 & 0.66   & 2.39 & 0.42 & 0.35  & 1.06 \\     \hline
B16-AGSS09met   \cite{vin}          &  6.03 & 4.50 &  1.46 & 4.50  & 8.25 & 2.04 & 1.44  & 3.26 \\ 
      $\mbox{\tiny errors}$           & 0.03 & 0.27 &  0.01 & 0.54  & 2.48 & 0.29 & 0.23  & 0.59 \\     \hline

\end{tabular}}
\end{table}

\noindent 
After the 
theory, 
we examine the data.
Pioneer observatories Homestake, Kamiokande, Gallex/GNO and SAGE 
saw solar neutrinos as reviewed by K.~Lande, T.~Kirsten, V.~Gavrin.\cite{dre} 
Their successors Super-Kamiokande and SNO measured precisely the 
B-neutrino flux, as discussed by Y.~Suzuki and H.~Robertson.\cite{dre}
Borexino determined accurately the  Be-neutrino component, 
studying B-neutrinos at the lowest energies and
probing  at $\sim$10\% the pp- and pep-fluxes  
as reported by M.~Wurm and D.~Guffanti.\cite{dre} 
Borexino's  bound on the sum of CNO-neutrinos, 
\begin{equation} \label{mab}
\color{blue}{\
\Phi_{\mbox{\tiny CNO}}\equiv \Phi_{\mbox{\tiny N}}+\Phi_{\mbox{\tiny N}}+ \Phi_{\mbox{\tiny F}} 
\equiv \varphi_{\mbox{\tiny CNO}} \times  \frac{10^{8}}{\mbox{cm$^2$s}}}
\end{equation}
is discussed later. To recap, 4 fluxes of the PP-chain out of 5 are known. 

\begin{figure}[t]
\begin{minipage}[c]{7.9cm}
\includegraphics[width=3.1in]{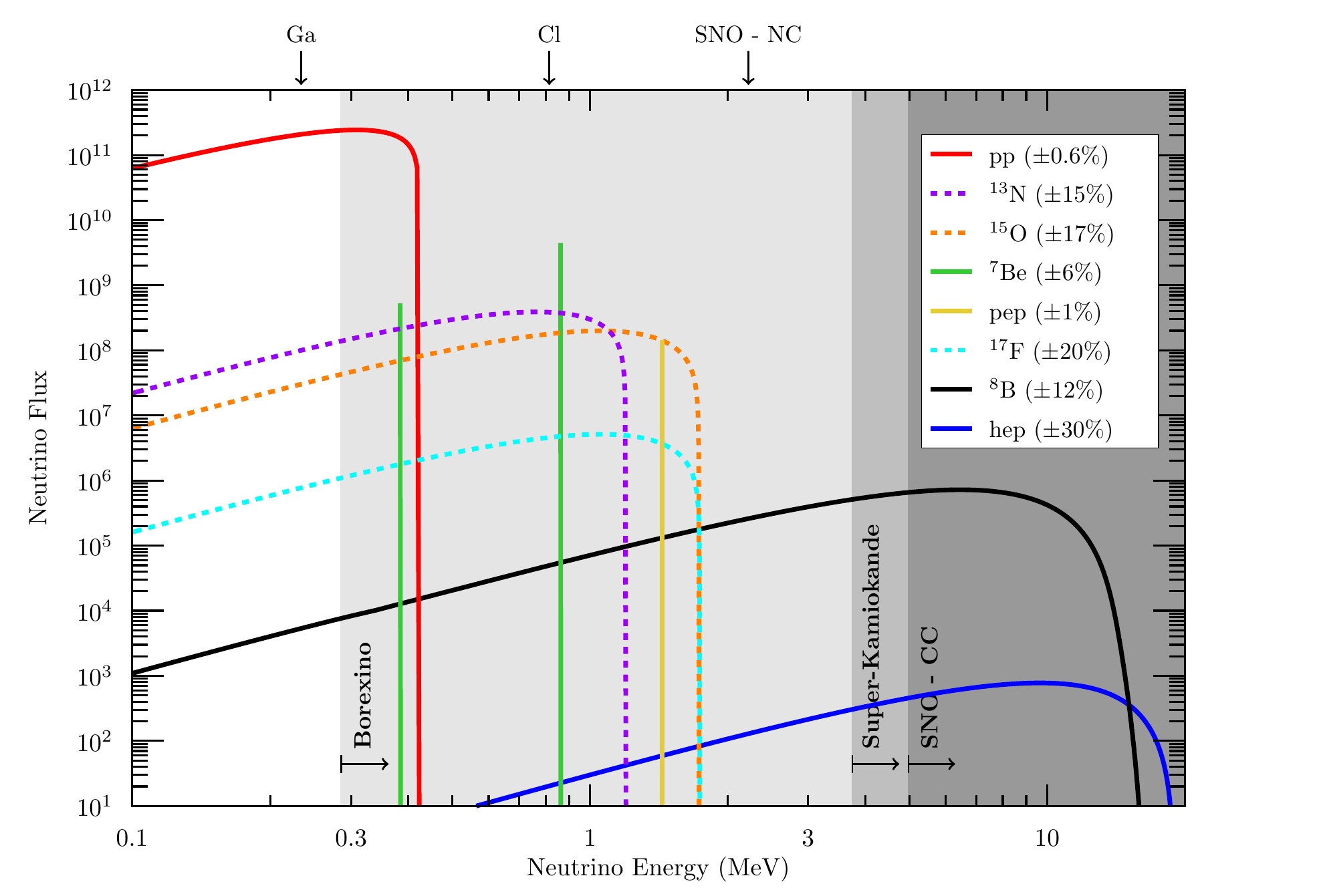}
\end{minipage}
\hskip1mm
\begin{minipage}[c]{3.2cm}
\caption{The 8 neutrino fluxes of the GS98 version of SSM, along with the 
sensitivity regions of the major present and past experiments.\cite{gr} The grey areas emphasize the regions of sensitivity of the few 
experiments that are 
able to probe the individual fluxes--see text for a discussion.
}
\end{minipage}
\label{fig1}
\end{figure}

\Fref{fig1} compares \textbf{SSM predictions} -- in 
the version that agrees with helioseismology as discussed by F.L.~Villante\cite{dre} --
and \textbf{energy thresholds} 
of various experiments, i.e., the minimum energy that they can probe.\cite{gr}
The experiments are grouped in 2 broad classes, i.e.,   \\
$1^{\mbox{\tiny st}}$ :-- the radiochemical experiments and the SNO-NC experiment.
They `integrate' above the energy threshold
indicated by the arrows given in the upper border of \fref{fig1},
or more precisely, they
measure a suitable average of the fluxes--see  
K.~Lande, T.~Kirsten, V.~Gavrin, H.~Robertson;\cite{dre} \\ 
$2^{\mbox{\tiny nd}}$ :--  the experiments 
based on elastic scattering on electrons $\nu+\mbox{e}\to \nu+\mbox{e}$ (ES)
and the SNO-CC experiment.
They probe the differential SSM fluxes in grey areas of \fref{fig1}, i.e., they discriminate the 
individual fluxes--see Y.~Suzuki,  
H.~Robertson, M.~Wurm and D.~Guffanti.\cite{dre} Note that,
\vskip-1.2mm
\noindent 
\begin{center}
\underline{the experiments sensitive to CNO-neutrinos are only SAGE and} 
\underline{Gallex/GNO, in the $1^{\mbox{\tiny st}}$ class, and Borexino,  in the $2^{\mbox{\tiny nd}}$ one.}
\end{center}
See\cite{vs1} for more discussion. 
Note that ES cross section is very well known.

\section{The luminosity constraint}\label{sec3}
In this section, we introduce and improve the description of 
the  ``luminosity constraint", that connects the neutrino fluxes to the observed solar luminosity.
The latter quantity--i.e., 
the
power emitted in light/heat--is measured much better than the neutrino fluxes,\cite{vin}  since the error
is below percent,
\begin{equation} \label{musu}
L_{\mathrel{\mathsmaller{\mathsmaller{\odot}}}}^{\mbox{\tiny obs}}=3.8418 \times 10^{33}\ \frac{\mbox{erg}}{\mbox{sec}}, \mbox{ with }
\frac{\delta L_{\mathrel{\mathsmaller{\mathsmaller{\odot}}}}^{\mbox{\tiny obs}}}{ L_{\mathrel{\mathsmaller{\mathsmaller{\odot}}}}^{\mbox{\tiny obs}}}
= 0.4\; \% 
\end{equation}
By definition, this is included in the SSM, but the use of the luminosity constraint is a milder assumption than the use of the SSM itself;
indeed, 
 solar neutrino data-analyses that implement it are called
{\em model independent.}

\subsection{Physical derivation of the luminosity constraint}
In the rudimentary  model of J.~Perrin,\cite{perro} 4 hydrogen atoms transform into 1  
helium atom, releasing energy   inside the Sun. Symbolically, 
\begin{equation}
4\mbox{H}\to {}^4\mbox{He}+ \mathcal{E}_{\mathrel{\mathsmaller{\mathsmaller{\odot}}}}
\end{equation}
The energy released, i.e., the difference between the masses 
of the initial and final atomic species (first measured by F.W.~Aston)  
is accurately known,  
\begin{equation}\label{epson}
\mathcal{E}_{\mathrel{\mathsmaller{\mathsmaller{\odot}}}}=4 M_1- M_4 \approx 26.73097\mbox{ MeV}
\end{equation} 
we use $c=1$. Today we know that  there are nuclear--not atomic--reactions in the Sun, but the estimation of the 
energy is almost unchanged. 
In fact, the effective nuclear reaction, $4p\to \alpha+2\nu_{\mbox{\tiny e}} +2\mbox{e}^+$, 
after adding 4 electrons on both sides, becomes equivalent to a reaction between atoms, 
\begin{equation} \label{memot}
4\mbox{H}\to {}^4\mbox{He}+ 2 \nu_{\mbox{\tiny e}} +  \varepsilon_{\mathrel{\mathsmaller{\mathsmaller{\odot}}}}
\mbox{ with } \varepsilon_{\mathrel{\mathsmaller{\mathsmaller{\odot}}}} \approx 26.1\mbox{ MeV} \
\end{equation}
We used the most recent version of SSM\cite{vin} to estimate the 
 energy lost in neutrinos, finding that it is small, and concluding that 
 $\varepsilon_{\mathrel{\mathsmaller{\mathsmaller{\odot}}}} \approx \mathcal{E}_{\mathrel{\mathsmaller{\mathsmaller{\odot}}}}$.

We begin with a  rough but useful estimation of 
the pp-neutrino flux (the main one) linked to the rate 
$\dot{N}_{\mbox{\tiny pp}}$ of the  $p+p\to D+ \nu_{\mbox{\tiny e}} +\mbox{e}^+ $ reaction
\begin{equation} 
\Phi_{\mbox{\tiny pp}}= \frac{ \dot{N}_{\mbox{\tiny pp}}  
}{4\pi\; \mbox{au}^2}
\end{equation} 
where au is the astronomical unit.  The rate can be estimated assuming 
$(i)$~that the solar luminosity is due to ${}^4\mbox{He}$ formation, 
$L_{\mathrel{\mathsmaller{\mathsmaller{\odot}}}}\approx \dot{N}_\alpha
\times \mathcal{E}_{\mathrel{\mathsmaller{\mathsmaller{\odot}}}}$
and $(ii)$~that helium nuclei are formed after 2 pp-reactions 
$ \dot{N}_\alpha \approx \dot{N}_{\mbox{\tiny pp}}/2 $.
From these positions,  $\Phi_{\mbox{\tiny pp}}$
turns out to be equal to the following flux, 
\begin{equation} \label{papar}
\color{blue}{\Phi_{\mathrel{\mathsmaller{\mathsmaller{\odot}}}}\equiv  \frac{2 L_{\mathrel{\mathsmaller{\mathsmaller{\odot}}}}}{4\pi\times \mbox{au}^2\times \mathcal{E}_{\mathrel{\mathsmaller{\mathsmaller{\odot}}}}} 
=6.379\times 10^{10}\frac{\nu_{\mbox{\tiny e}}}{\mbox{cm$^2$ s}} (1\pm 0.4\%)}
\end{equation}
While the result $\Phi_{\mbox{\tiny pp}} \approx \Phi_{\mathrel{\mathsmaller{\mathsmaller{\odot}}}}$ agrees within $\sim 10$\% with SSM,  the above derivation  neglects the energy lost in neutrinos and, more importantly, it does not consider the presence of any other neutrinos except pp's.

\begin{sidewaystable}[p]
\tbl{\small 1$^{\mbox{\tiny st}}$ column;  sequence of nuclear reactions. 
2$^{\mbox{\tiny nd}}$ column;  identification code of the sequence.
3$^{\mbox{\tiny rd}}$ column; $\nu_{\mbox{\tiny e}}$ flux associated to the sequence.
4$^{\mbox{\tiny th}}$ column; mass difference $\Delta M$ of the  sequence.$^\dagger$
5$^{\mbox{\tiny th}}$ column; energy loss in neutrino $\langle E_\nu \rangle$. 
6$^{\mbox{\tiny th}}$ column; amount of light/heat $\varepsilon_i$ in the sequence.
The rows are grouped as follows; 2$^{\mbox{\tiny nd}}$ and 3$^{\mbox{\tiny rd}}$, sequences of reactions of the PP-chain where ${}^3$He is formed; 4$^{\mbox{\tiny th}}$-7$^{\mbox{\tiny th}}$, 
sequences of  
reactions of the PP-chain where ${}^3$He is consumed; 8$^{\mbox{\tiny th}}$ and 9$^{\mbox{\tiny th}}$, 
sequences of reactions of the CNO-I-cycle; 10$^{\mbox{\tiny th}}$ and 11$^{\mbox{\tiny th}}$, sequences of CNO-II.
\label{tab-33}}
{\begin{tabular}{l c c | c  c c }
\hline
sequences & sequence & asso- & relevant mass  & $\nu_{\mbox{\tiny e}}$ loss & light/   \\ 
of nuclear & code & ciated  &  difference $\Delta M_a$ & $\langle E_{\nu }\rangle$  & heat $\varepsilon_a$    \\ 
reactions & $a$ &  $\nu_{\mbox{\tiny e}}$ flux 
&   (atomic species) &   [MeV] & [MeV]    \\ 
\hline \hline\\[-1ex]
 $ \mbox{\small p}(\mbox{\small p},\mbox{\small e}^{\mbox{\tiny +}}\nu_{\mbox{\tiny e}})\mbox{\small d}(\mbox{\small p},\gamma){}^3\mbox{\small He}$  &
  $p1$ & $\Phi_{\mbox{\tiny pp}}$ & 
 $\scriptstyle 3M_1-M_3$ &0.2668 &  6.6689 \\[1ex]
 $   \mbox{\small p}(\mbox{\small p}\ \mbox{\small e}^{\mbox{\tiny -}},\nu_{\mbox{\tiny e}})\mbox{\small d}(\mbox{\small p},\gamma){}^3\mbox{\small He}$ &
  $pe1$ &  $\Phi_{\mbox{\tiny pep}}$ &   
 $\scriptstyle 3M_1-M_3$ &1.445 &  5.491 \\[1ex] \hline \\[-1ex]
$ {}^3\mbox{\small He}({}^3\mbox{\small He}, 2 \mbox{\small p}){}^4\mbox{\small He}$ &
33 & $\Phi_{33}$ & 
$\scriptstyle 2M_3-M_4-2M_1$ &0.000& 12.8596\\[1ex]  
 $ {}^3\mbox{\small He}({}^4\mbox{\small He}, \gamma){}^7\mbox{\small Be}(\mbox{\small e}^{\mbox{\tiny -}},\nu_{\mbox{\tiny e}}){}^7\mbox{\small Li}(\mbox{\small p},{}^4\mbox{\small He}){}^4\mbox{\small He} $ &
 $e7$ & $\Phi_{\mbox{\tiny Be}}$ & 
 $\scriptstyle M_3 + M_1-M_4$ &0.813& 18.982 \\[1ex]  
 $ {}^3\mbox{\small He}({}^4\mbox{\small He}, \gamma){}^7\mbox{\small Be}(\mbox{\small p},\gamma){}^8\mbox{\small B}(\mbox{\small e}^{\mbox{\tiny +}}\nu_{\mbox{\tiny e}}){}^8\mbox{\small Be}^*({}^4\mbox{\small He}) {}^4\mbox{\small He} $ &
 $p7$ & $\Phi_{\mbox{\tiny B}}$ & 
$\scriptstyle M_3 + M_1-M_4$ &6.735& 13.060 \\[1ex]  
 $ {}^3\mbox{\small He}(\mbox{\small p} , \mbox{\small e}^{\mbox{\tiny +}}\nu_{\mbox{\tiny e}}){}^4\mbox{\small He}$ &
 $p3$ & $\Phi_{\mbox{\tiny hep}}$ & 
$\scriptstyle M_3 + M_1-M_4$ &9.630&  10.165 \\[1ex] \hline\hline \\[-1ex]
$ {}^{12}\mbox{\small C}(\mbox{\small p} , \gamma){}^{13}\mbox{\small N}(\mbox{\small e}^{\mbox{\tiny +}}\nu_{\mbox{\tiny e}}){}^{13}\mbox{\small C}(\mbox{\small p},\gamma){}^{14}\mbox{\small N}$ &
 $p12$  &   $\Phi_{\mbox{\tiny N}}$ & 
$\scriptstyle M_{12} + 2 M_1-M_{14}$ &0.707& 11.008\\[1ex]
 $ {}^{14}\mbox{\small N}(\mbox{\small p} , \gamma){}^{15}\mbox{\small O}(\mbox{\small e}^{\mbox{\tiny +}}\nu_{\mbox{\tiny e}}){}^{15}\mbox{\small N}(\mbox{\small p},{}^{12}\mbox{\small C}){}^4\mbox{\small He}$ &
 $p14$ & $\Phi^{\mbox{\tiny I}}_{\mbox{\tiny O}}$ & 
$\scriptstyle M_{14} + 2 M_1-M_{12}-M_4$  &0.997& 14.019 \\[1ex] \hline \\[-1ex]
 $ {}^{14}\mbox{\small N}(\mbox{\small p} , \gamma){}^{15}\mbox{\small O}(\mbox{\small e}^{\mbox{\tiny +}}\nu_{\mbox{\tiny e}}){}^{15}\mbox{\small N}
 $ &
 $p14'$ & $\Phi^{\mbox{\tiny II}}_{\mbox{\tiny O}}$ & 
$\scriptstyle M_{14} + M_1-M_{15}$  &0.997& 9.054 \\[1ex] 
 $ {}^{15}\mbox{\small N}(\mbox{\small p} , \gamma){}^{16}\mbox{\small O}(\mbox{\small p} , \gamma){}^{17}\mbox{\small F}(\mbox{\small e}^{\mbox{\tiny +}}\nu_{\mbox{\tiny e}})
{}^{17}\mbox{\small O}(\mbox{\small p} {}^{14}\mbox{\small N}){}^4\mbox{\small He} $ &
 $p15$ & $\Phi_{\mbox{\tiny F}}$ & 
$\scriptstyle M_{15} + 3 M_1-M_{14}-M_4$  &1.000& 15.680  \\[1ex] \hline 
\multicolumn{6}{c}{$^\dagger$Note 
the relations 
$\mathcal{E}_{\mathrel{\mathsmaller{\mathsmaller{\odot}}}}\equiv 4M_1-M_4=
\Delta M_{{\tiny p1}} + \Delta M_{{\tiny p7}}=2 \times \Delta M_{{\tiny p1}}+ \Delta M_{{\tiny 33}}=
\Delta M_{{\tiny p12}} + \Delta M_{{\tiny p14}}=
\Delta M_{{\tiny p14'}} + \Delta M_{{\tiny p15}}$ etc.
}
\end{tabular}
}
\end{sidewaystable}

\begin{figure}[t]
\begin{minipage}[c]{7.cm}
\includegraphics[width=2.8in]{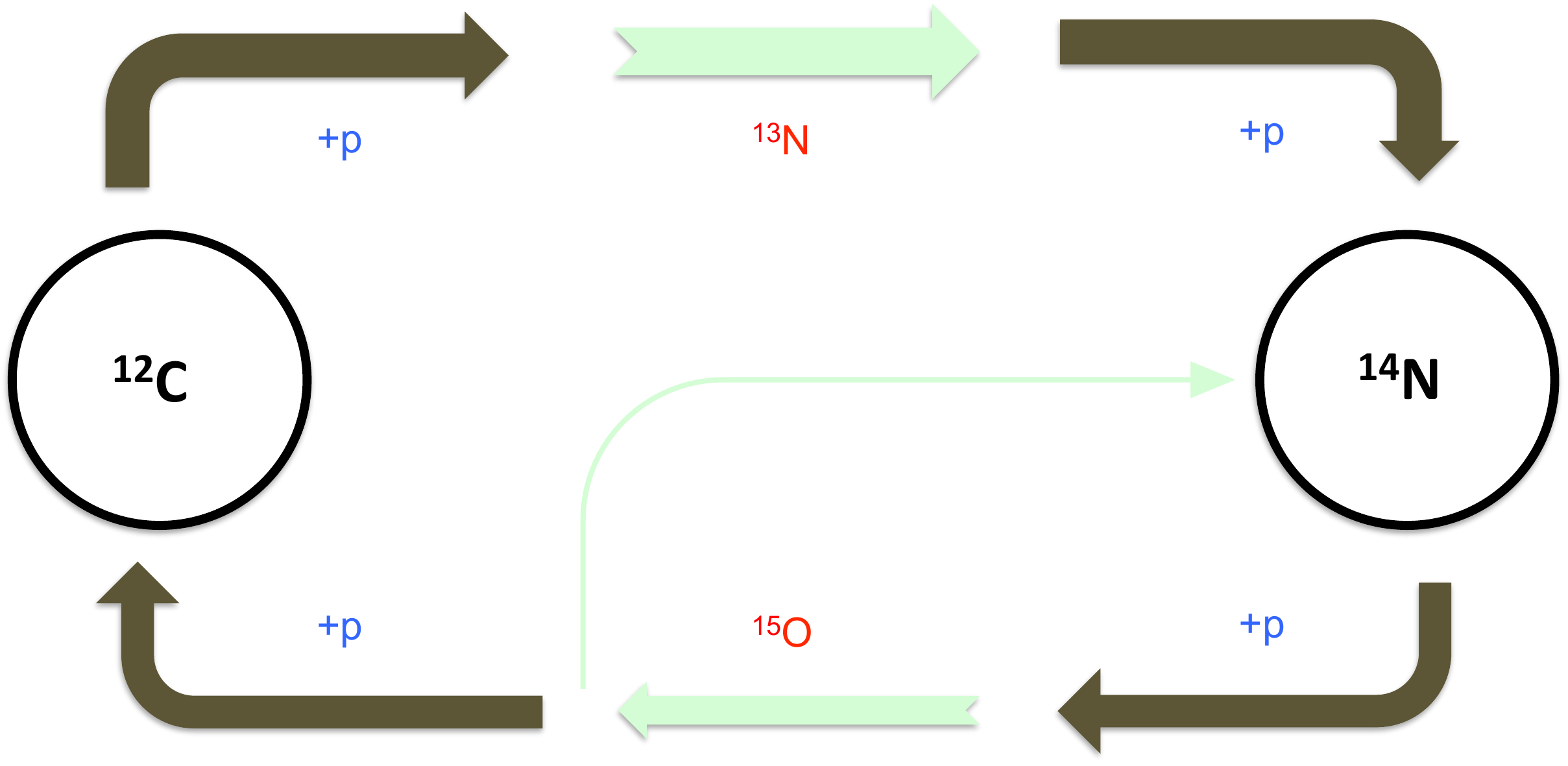}
\end{minipage}
\hskip1mm
\begin{minipage}[c]{4.1cm}
\caption{Scheme of the reactions of the
CNO-I cycle.
The light green arrows indicate the $\beta^+$ decay reactions 
that are tagged by neutrinos; the dark green arrows those that are not. 
The size of the arrows, not in scale, indicates the probability of the specific reactions.
(The small internal arrow recalls us the existence of the CNO-II cycle.)
}
\label{fig2}
\end{minipage}
\end{figure}

Both these shortcomings can be easily made up by positing,
\begin{equation} \label{citat0}
{L_{\mathrel{\mathsmaller{\mathsmaller{\odot}}}}}  + 
L_{\nu_{\mbox{\tiny e}}}
=
\mathcal{E}_{\mathrel{\mathsmaller{\mathsmaller{\odot}}}} \times \dot{N}_\alpha
\end{equation}
The term on   r.h.s.\ corresponds to 
the energy input (the source). \Eref{memot}  prescribes that, when 
1 $^4$He is produced, 2 neutrinos are released, 
thus, 
\begin{equation} \label{clitoi0}
 \mathcal{E}_{\mathrel{\mathsmaller{\mathsmaller{\odot}}}} \times 
\dot{N}_\alpha= 
\mathcal{E}_{\mathrel{\mathsmaller{\mathsmaller{\odot}}}}
\times 
\frac{\dot{N}_\nu}{2}
\mbox{ with }
 \dot{N}_\nu\equiv 4\pi\; \mbox{au}^2\times \sum_i \Phi_i
\end{equation} 
where $\dot{N}_\nu$ is 
the number of neutrinos emitted by the Sun per second. 
The l.h.s\ terms 
are the energy outputs, `gains' and `losses' (i.e., the fluxes due to light and to 
neutrinos); the `gains' are measured, the `losses' are, 
\begin{equation} \label{citatw0}
L_{\nu_{\mbox{\tiny e}}} = 4\pi\; \mbox{au}^2 \times 
\sum_i \langle E_i \rangle\ \Phi_i
\end{equation}
Eqs.~(\ref{citat0}-\ref{citatw0}) boil down to a constraint for the neutrino fluxes--namely, the luminosity 
constraint, that using \eref{papar}, has the compact expression,\cite{2001}
\begin{equation} \label{clitoi00}
 \sum_i \Phi_i \left( 1 -    \frac{2 \langle E_i \rangle\ }{\mathcal{E}_{\mathrel{\mathsmaller{\mathsmaller{\odot}}}}} \right)
 =\Phi_{\mathrel{\mathsmaller{\mathsmaller{\odot}}}}\ ,\ 
 \mbox{ where }
{i=\mbox{\small pp,Be,pep,B,hep,N,O,F}}
\end{equation} 
The values of the average  energies $\langle E_i \rangle$ are given in \tref{tab-33}.
The  
constraint based on \eref{clitoi00}
is already an excellent approximation. 
Its accuracy can be further enhanced describing the possibility that the process of 
synthesis of $^4$He from the CNO-cycle 
is incomplete. In order to do so, we consider \eref{citat0}, and
after dividing  
by $4 \pi\times \mbox{au}^2$, we include one last term,
\begin{equation} \label{cita1}
\frac{L_{\mathrel{\mathsmaller{\mathsmaller{\odot}}}}}{4 \pi\times \mbox{au}^2}   + \sum_i \langle E_i \rangle\ \Phi_i
=
\frac{\mathcal{E}_{\mathrel{\mathsmaller{\mathsmaller{\odot}}}}}{2}\  \sum_i \Phi_i -
\left( \frac{\mathcal{E}_{\mathrel{\mathsmaller{\mathsmaller{\odot}}}}}{2}-  {\mathcal{E}}_{ \mbox{\tiny C$\to$N}} \right)  \delta\Phi_{ \mbox{\tiny CNO}} 
 \end{equation}
This last term appears as follows:
The CNO-I cycle stops after the production of $^{14}$N  in a significant number of cases 
 (upper  branch of \fref{fig2}) and this  
leads to an excess of $\Phi_{ \mbox{\tiny N}}$ over $\Phi_{ \mbox{\tiny O}}$.
The corresponding energy release is $4\pi\; \mbox{au}^2$ times 
${\mathcal{E}}_{ \mbox{\tiny C$\to$N}}\Phi_{ \mbox{\tiny N}} 
+ ({\mathcal{E}_{\mathrel{\mathsmaller{\mathsmaller{\odot}}}}}-{\mathcal{E}}_{ \mbox{\tiny C$\to$N}}) \Phi_{ \mbox{\tiny O}}$
that can be cast as on the r.h.s.\ of \eref{cita1} with
$\delta\Phi_{ \mbox{\tiny CNO}}= \Phi_{ \mbox{\tiny N}}   -\Phi_{ \mbox{\tiny O}}$. 
The CNO-II cycle,  if assumed to be at kinetic equilibrium,
leads to a further energy release of 
$4\pi\; \mbox{au}^2\times {\mathcal{E}_{\mathrel{\mathsmaller{\mathsmaller{\odot}}}}} \Phi_{ \mbox{\tiny F}}$.
If we replace  
${\mathcal{E}}_{ \mbox{\tiny C$\to$N}}\Phi_{ \mbox{\tiny N}} 
+ ({\mathcal{E}_{\mathrel{\mathsmaller{\mathsmaller{\odot}}}}}-{\mathcal{E}}_{ \mbox{\tiny C$\to$N}}) \Phi_{ \mbox{\tiny O}}$
with 
${\mathcal{E}}_{ \mbox{\tiny C$\to$N}}\Phi_{ \mbox{\tiny N}} 
+ ({\mathcal{E}_{\mathrel{\mathsmaller{\mathsmaller{\odot}}}}}-{\mathcal{E}}_{ \mbox{\tiny C$\to$N}}) 
(\Phi_{ \mbox{\tiny O}}-\Phi_{ \mbox{\tiny F}})+
{\mathcal{E}_{\mathrel{\mathsmaller{\mathsmaller{\odot}}}}} \Phi_{ \mbox{\tiny F}}$, see \fref{fig2},
this can be cast as on the r.h.s.\ of \eref{cita1} defining
\begin{equation} \label{clita}
\color{blue}{\delta\Phi_{ \mbox{\tiny CNO}}= \Phi_{ \mbox{\tiny N}}  + \Phi_{ \mbox{\tiny F}}
-\Phi_{ \mbox{\tiny O}}}
\end{equation} 
The numerical value of the energy of synthesis of 
$^{14}$N  from $^{12}$C (after the absorption of two protons)
is given in $4^{ \mbox{\tiny th}}$ column,   
$8^{ \mbox{\tiny th}}$ row of of \tref{tab-33},
\begin{equation} \label{papibot}
\mathcal{E}_{ \mbox{\tiny C$\to$N}} =M_{12}+2 M_1-M_{14}=11.715\mbox{ MeV}
\end{equation} 
This is similar to ${\mathcal{E}_{\mathrel{\mathsmaller{\mathsmaller{\odot}}}}}/2$: thus, the last term of \eref{cita1} is suppressed.
In the next section, a different way to arrive at \eref{cita1} is described.

\subsection{Another derivation  of the luminosity constraint} 
Bahcall's derivation of the luminosity constraint 
begins from the position,\cite{lum}
\begin{equation} \label{lamantino}
L_{\mathrel{\mathsmaller{\mathsmaller{\odot}}}}  =4 \pi\times \mbox{au}^2 \times 
\sum_{i} \mathcal{Q}_i\times \Phi_i
\end{equation}
The coefficients 
$\mathcal{Q}_i$ are the amount of light/heat 
associated to each of the 8 neutrino fluxes of the SSM, $\Phi_i$.
%
We will show that a formal derivation of the coefficients $\mathcal{Q}_i$ leads  again to \eref{cita1}--see
 Eqs.~(\ref{cciefa},\ref{cciefa2},\ref{kuly}-\ref{citabo}) later on.



\begin{figure}[t]
\begin{center}
\includegraphics[width=3.4in]{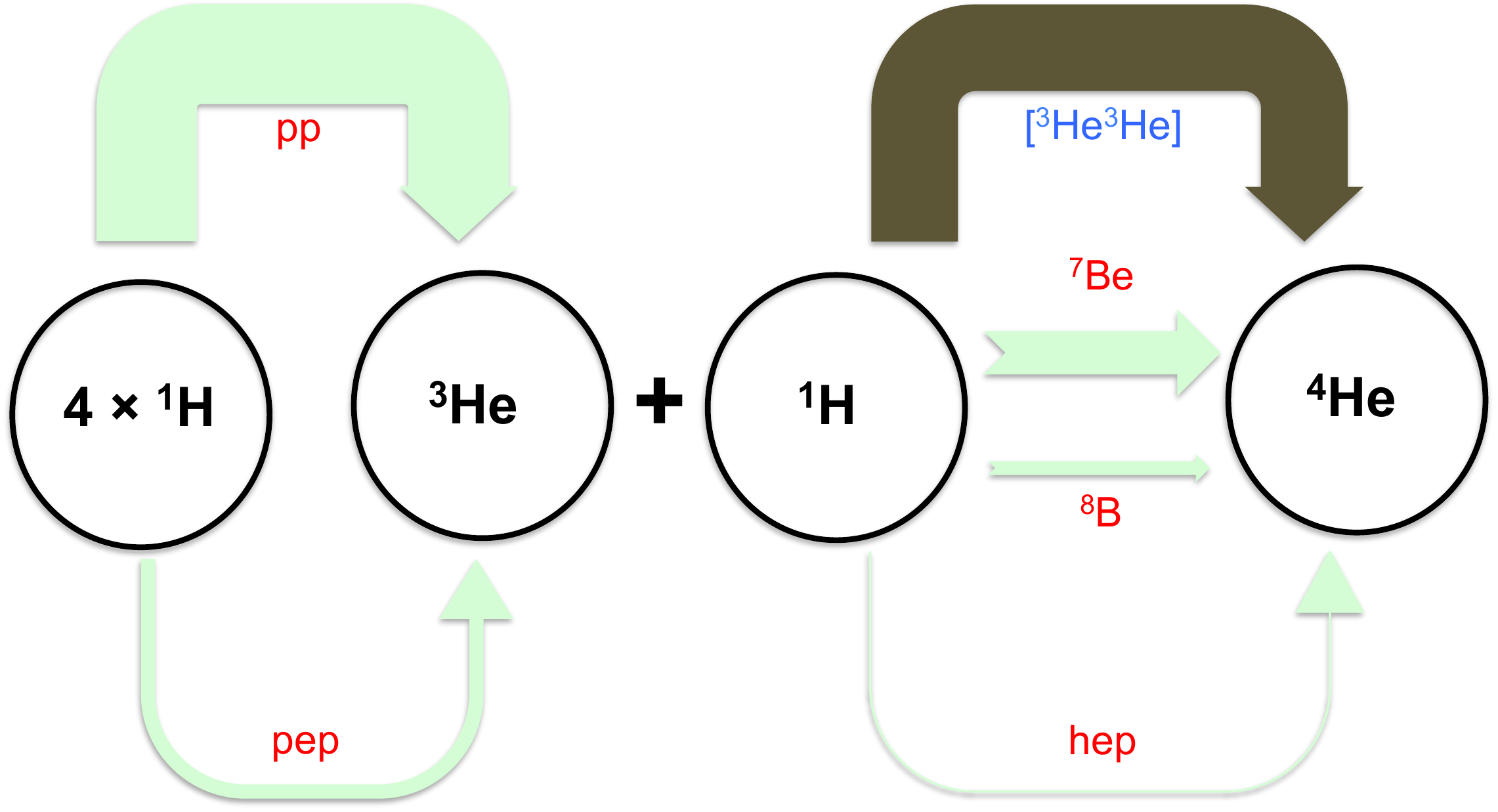}
\end{center}
\caption{A representation of the sequences of the PP-chain in 
\tref{tab-33}. 
The light green arrows indicate the sequences of reactions 
that are tagged by the 5 neutrino fluxes while the dark green arrow indicates the single sequence that is not 
tagged by neutrinos. 
Note the natural division of the PP-chain in the $^3$He formation (first part, fully tagged by pp and pep-neutrinos)
and $^3$He consumption (second part, tagged only partly).
The size of the arrows (not in scale) remind us the different probabilities of the sequences.
}
\label{fig3}
\end{figure}

The reactions of the PP-chain and of the CNO-cycles, that allow the Sun to shine,  
can be grouped in  10 {\em sequences}, see  \tref{tab-33}.
The PP-chain is divided in 6 sequences: 2 that form ${}^3$He and 4 that consume it, 
see \fref{fig3} for illustration. In the Sun, the CNO-cycle has two active sub-cycles, the CNO-I and the CNO-II; 
there are 2 sequences in each of them--those of the CNO-I sub-cycle are the upper and lower branches of \fref{fig2}.

The values of the index $a$ 
that identify the 10 sequences 
are given in the
2$^{\mbox{\tiny nd}}$ column of \tref{tab-33}.
Each  sequence comprises various reactions
(given in the $1^{\mbox{\tiny st}}$ column of \tref{tab-33})  
that are in kinetic equilibrium among them--i.e., all reactions have the same rate.
Most sequences--8, out of 10--are associated to one specific neutrino flux: this allows us to {\bf measure the reaction rate $\dot{N}_a$}
of each sequence, that is  given by, 
\begin{equation} \label{defa}
\color{blue}{{\dot{N}_a }= {4\pi\times \mbox{au}^2}\times \Phi_a}
\end{equation}
Consider now  
the amount of light/heat $\varepsilon_a$
produced in each sequence.
The luminosity, usually measured in erg/s, 
can be written as, 
\begin{equation} \label{lefa}
L_{\mathrel{\mathsmaller{\mathsmaller{\odot}}}} = \sum_a \varepsilon_a \times \dot{N}_a 
\end{equation}
The values of $\varepsilon_a$ are given 
in the last column of \tref{tab-33} and are obtained as follows.
The initial and final species of a  
sequence of reactions lead to energy release. This  
goes partly in kinetic energy or positrons 
(thus contributing to light/heat) and partly in the 
neutrino (being then lost):
\begin{equation} \label{imelda}
\Delta M_a  = \varepsilon_a +  \langle E_{\nu}\rangle_{\! a}
\end{equation}
where $\Delta M=M_{\mbox{\tiny initial}}-M_{\mbox{\tiny final}}$ is the difference of atomic masses in the sequence, 
given 4$^{\mbox{\tiny th}}$ column of \tref{tab-33}; the average energy losses  $\langle E_{\nu}\rangle$
(5$^{\mbox{\tiny th}}$ column of \tref{tab-33})
are as in Bahcall.\cite{jb,lum,bahlu} Now we consider the two cases:


\noindent 
{\em PP-chain:}  As emphasized in \fref{fig3},
no physical neutrino flux is produced in the sequence 33.
Still, this sequence can be associated to one  formal neutrino flux, assuming with 
Bahcall\cite{lum} that the rate of formation of $^3$He in the PP-chain,  $\dot{N}_{p1}+ \dot{N}_{pe1}$, 
  equates the rate of consumption 
of $^3$He, $2\times \dot{N}_{33}+ \dot{N}_{e7}+\dot{N}_{p7}+ \dot{N}_{p3}$. In other words, we extend \eref{defa} introducing
\begin{equation}\label{33fa}
\Phi_{33} \equiv \frac{\dot{N}_{33}}{ 4\pi\; \mbox{au}^2}
=\frac{1}{2} ( \Phi_{ \mbox{\tiny pp}} +\Phi_{ \mbox{\tiny pep}}
-\Phi_{ \mbox{\tiny Be}} - \Phi_{ \mbox{\tiny B}} -\Phi_{ \mbox{\tiny hep}}    )
\end{equation}
This formal neutrino flux
{\em does not entail} energy losses and   
can be evaluated from the observable fluxes of the PP-chain.
Using  Eqs.~(\ref{defa},\ref{imelda},\ref{33fa}), we rearrange
the first 6 terms of \eref{lefa}   as in \eref{lamantino} 
finding, 
\begin{equation} \label{cciefa}
\begin{array}{l}
\mathcal{Q}_{\mbox{\tiny pp}} = \varepsilon_{p1} +  \frac{\varepsilon_{33}}{2}=\frac{\mathcal{E}_{\mathrel{\mathsmaller{\mathsmaller{\odot}}}}}{2} - \langle E_{\mbox{\tiny pp}} \rangle \ ,\  
\mathcal{Q}_{\mbox{\tiny pep}} = \varepsilon_{pe1} +  \frac{\varepsilon_{33}}{2}=\frac{\mathcal{E}_{\mathrel{\mathsmaller{\mathsmaller{\odot}}}}}{2} - \langle E_{\mbox{\tiny pep}} \rangle, 
 \\[0.3ex]
\mathcal{Q}_{\mbox{\tiny Be}} = \varepsilon_{p7}  -  \frac{\varepsilon_{33}}{2}=\frac{\mathcal{E}_{\mathrel{\mathsmaller{\mathsmaller{\odot}}}}}{2} - \langle E_{\mbox{\tiny Be}} \rangle
\ ,\  
\mathcal{Q}_{\mbox{\tiny B}} = \varepsilon_{e7}  -  \frac{\varepsilon_{33}}{2}=\frac{\mathcal{E}_{\mathrel{\mathsmaller{\mathsmaller{\odot}}}}}{2} - \langle E_{\mbox{\tiny B}} \rangle, 
\\[0.3ex]
\mathcal{Q}_{\mbox{\tiny hep}} = \varepsilon_{p3}  -  \frac{\varepsilon_{33}}{2}=\frac{\mathcal{E}_{\mathrel{\mathsmaller{\mathsmaller{\odot}}}}}{2} - \langle E_{\mbox{\tiny hep}} \rangle,
 \end{array}
 \end{equation}
where $\mathcal{E}_{\mathrel{\mathsmaller{\mathsmaller{\odot}}}}$ is as in 
\eref{epson} and the other quantities are as in \tref{tab-33}.

\noindent 
{\em CNO-cycle:}  The O-neutrinos track mostly the CNO-I cycle
and only in a minor amount the CNO-II cycle:  $\Phi_{ \mbox{\tiny O }}^{ \mbox{\tiny I }}  \gg \Phi_{ \mbox{\tiny O }}^{ \mbox{\tiny II }}$.   
In terms of the coefficients given in \tref{tab-33}, the 
contribution to the luminosity of the CNO-cycle reads,
\begin{equation}
\frac{L_{\mathrel{\mathsmaller{\mathsmaller{\odot}}}}^{ \mbox{\tiny CNO }}}{4 \pi\times \mbox{au}^2}  =
\varepsilon_{p12}\; \Phi_{ \mbox{\tiny N }} + 
\varepsilon_{p14}\; \Phi_{ \mbox{\tiny O }}^{ \mbox{\tiny I }}  + 
\varepsilon_{p14'}\; \Phi_{ \mbox{\tiny O }}^{ \mbox{\tiny II }}  + 
\varepsilon_{p15}\; \Phi_{ \mbox{\tiny F }} 
\end{equation}
Assume kinetic equilibrium for CNO-II, 
$\Phi_{ \mbox{\tiny O}}=\Phi_{ \mbox{\tiny O}}^{ \mbox{\tiny I}}
+\Phi_{ \mbox{\tiny O}}^{ \mbox{\tiny II}}$
and
$\Phi_{ \mbox{\tiny F}} = 
\Phi_{ \mbox{\tiny O}}^{ \mbox{\tiny II}}$; then, 
inserting the expressions for $\varepsilon_a$ from  
\tref{tab-33} and  \eref{imelda}, this becomes, 
\begin{equation} \label{pefa}
\frac{L_{\mathrel{\mathsmaller{\mathsmaller{\odot}}}}^{ \mbox{\tiny CNO }}}{4 \pi\times \mbox{au}^2}  =
\varepsilon_{p12}\; \Phi_{ \mbox{\tiny N }} + 
\varepsilon_{p14}\; \Phi_{ \mbox{\tiny O }} + 
(\varepsilon_{p12} +  \langle E_{\mbox{\tiny N} } \rangle -  \langle E_{\mbox{\tiny F} }\rangle  )\; \Phi_{ \mbox{\tiny F }} 
\end{equation}
Proceeding as for the PP chain and using Eqs.~(\ref{imelda},\ref{pefa}), 
we reorganize  the 4 terms of the 
sum of \eref{lefa} as in \eref{lamantino} finding
\begin{equation} \label{cciefa2}
\begin{array}{l}
 \mathcal{Q}_{\mbox{\tiny N}} = \varepsilon_{p12}  = {\mathcal{E}_{\mbox{\tiny C$\to$N}}} - \langle E_{\mbox{\tiny N}} \rangle
  \ ,\   
  \mathcal{Q}_{\mbox{\tiny O}} = \varepsilon_{p14}  = {\mathcal{E}_{\mathrel{\mathsmaller{\mathsmaller{\odot}}}}} - {\mathcal{E}_{\mbox{\tiny C$\to$N}}} - \langle E_{\mbox{\tiny O}} \rangle,
 \\
\mathcal{Q}_{\mbox{\tiny F}} = \varepsilon_{p12}  +  \langle E_{\mbox{\tiny N} } \rangle -  \langle E_{\mbox{\tiny F} }\rangle 
= {\mathcal{E}_{\mbox{\tiny C$\to$N}}} - \langle E_{\mbox{\tiny F}} \rangle,
%
\end{array}
\end{equation}
where $\mathcal{E}_{ \mbox{\tiny C$\to$N}}$ is as in \eref{papibot}.
If one wants to 
omit the contribution of the 
CNO-II chain, describing the assumption that this is fully out of 
kinetic equilibrium  (as Bahcall did\cite{lum}),  
it is enough to set $\Phi_{ \mbox{\tiny F}}=0$.



\begin{table}[t]
\tbl{Coefficients, in MeV, to express $L_{\mathrel{\mathsmaller{\mathsmaller{\odot}}}}$  as a function of the observable 
neutrino fluxes.\label{tabend}}
{\begin{tabular}{ c | ccccc | ccc }
&\multicolumn{5}{c}{The PP-chain} &\multicolumn{3}{c}{The CNO-cycle}\\
\toprule
 & ${\mbox{\small pp}}$ & ${\mbox{\small Be}}$ &  ${\mbox{\small pep}}$ &  
${\mbox{\small B}}$  &  ${\mbox{\small hep}}$ &   ${\mbox{\small N}}$ &  ${\mbox{\small O}}$ & ${\mbox{\small F}}$  \\[1ex] 
\toprule
$\mathcal{Q}_i^{\mbox{\tiny this work}}$  & 
$\scriptstyle 13.0987$ & $\scriptstyle 12.5525$ & $\scriptstyle 11.9205$ &  $\scriptstyle 6.6305$ & $\scriptstyle 3.7355$ & 
$\scriptstyle 11.0075$ &$\scriptstyle 14.0194$ &  $\scriptstyle 10.715$ \\ \hline
$\mathcal{Q}_i^{\mbox{\tiny Bahcall}}$  & 
$\scriptstyle 13.0987$ & $\scriptstyle 12.6008$ &$\scriptstyle  11.9193$ & $\scriptstyle 6.6305$ & $\scriptstyle 3.7370$ & 
$\scriptstyle 3.4577$ & $\scriptstyle 21.5706$ &  $\scriptstyle 0$ \\
\hline \hline 
$\Delta\mathcal{Q}_i\times 1000$  & 
$\scriptstyle 0$ & $\scriptstyle 48.3$ &$\scriptstyle  -1.2$ & $\scriptstyle 0$ & $\scriptstyle 1.5$ & 
$\scriptstyle -7549.8$ &$\scriptstyle 7551.2$ &  $\scriptstyle -10715$ \\
\hline
\end{tabular}}
\end{table}

\subsection{Improved values of the luminosity coefficients}
At this point we can summarize. Let us begin recalling \eref{lamantino}, namely,
\begin{equation} \label{kuly}
L_{\mathrel{\mathsmaller{\mathsmaller{\odot}}}}  =4 \pi\times \mbox{au}^2 \times 
\sum_{i} \mathcal{Q}_i\times \Phi_i 
\mbox{ with }
{i=\mbox{\small pp,Be,pep,B,hep,N,O,F}}
\end{equation}
\vskip-3mm
\noindent where,
\begin{equation}  \label{kuly2}
\left\{
\begin{array}{lr}
\mathcal{Q}_i=\mbox{light/heat associated to the $i$-th flux} &  \mbox{[\ {\small MeV}\ ]}\\[-0.2ex]
\Phi_i=\mbox{observable solar neutrino fluxes}  & \mbox{[\ {\small 1/(cm$^2$s)}\ ]}\\[-0.2ex]
4 \pi\times \mbox{au}^2=4.50579\times 10^{21} &  {[\ \mbox{\small erg cm}^2}/{\mbox{\small MeV}\ ]} 
\end{array}
\right.
\end{equation}
The numerical values of the coefficients $\mathcal{Q}_i$ 
calculated with  \tref{tab-33}  and 
Eqs.~(\ref{cciefa},\ref{cciefa2}) 
are given in \tref{tabend}: They are perfectly compatible 
with the outcomes of the SSM.  
In the same table one finds the values obtained by Bahcall.\cite{lum}
Comparing the two sets of values, we note that,

{\footnotesize
\begin{enumerate}
\item Two contributions are in perfect agreement,
those of pp- and B-fluxes.
\item Two differences are of trivial interpretation, namely, those of 
the pep- and hep-contributions, that are small and within roundoff errors.
\item The F-flux contribution to luminosity was simply omitted by Bahcall.\footnote{ 
This contribution  to the luminosity is expected to be as or more important 
than B-flux one in the SSM; anyway it is small and can be neglected for current accuracy.}
\item Some differences are non-trivial and need discussion: those of Be, N, O.
\end{enumerate}}

\noindent 
{\em Treatment of  Be-neutrinos:} 
Two beryllium lines cause energy losses. The treatment advocated by Bahcall\cite{lum} requires two steps:
\begin{quote}
{\footnotesize\it ``one must average over the two $^7$Be neutrino lines with the appropriate weighting \underline{and} include the $\gamma$-ray energy from the 10.3\% of the decays that go to the first excited state of $^7$Li.''

}
\end{quote}
(the emphasis on  the word ``{and}'' is our choice and not in the original text). 

\vskip3mm
\noindent In our understanding, Bahcall's procedure is incorrect, for,
{\footnotesize
\begin{enumerate}
\item The energy loss of the EC on $^7$Be is due to a transition to the ground state in $\sim 90$\% the cases, 
and to the one in the $1^{\mbox{\tiny st}}$ excited level in the rest; the average 
loss is thus $0.895 \times 863.1 + 0.105 \times 385.5 = 813.0$ keV as quoted in \tref{tab-33}
(we updated $10.3 \to 10.5$\%, following Tilley et al 2002,\cite{till} even though this is not crucial). 
\item The gain of energy due to the $\gamma$-ray of 477.61 keV, emitted in 10.5\% of the cases by the first excited state of $^7$Li, amounts to 50.1 keV. However, we do not include it as Bahcall does, for this amounts to {\bf double counting:} to be sure, $863.1 -  813.0 = 50.1.$ 
\end{enumerate}}

\noindent 
{\em Treatment of  N- and O-neutrinos:} 
The difference with Bahcall is in the choice of the sequences of the CNO-I cycle.\footnote{
The $1^{\mbox{\tiny st}}$ sequence 
according to him is    
$ {}^{12}\mbox{\small C}(\mbox{\small p} , \gamma){}^{13}\mbox{\small N}(\mbox{\small e}^{\mbox{\tiny +}}\nu_{\mbox{\tiny e}}){}^{13}\mbox{\small C}$:
this associates only very little energy  to the release of N-neutrinos - and conversely much more to O-neutrinos.
This choice differs from the one of \tref{tab-33}; instead, 
the values of $\mathcal{Q}_ {\mbox{\tiny N}} + \mathcal{Q}_ {\mbox{\tiny O}}$  agree well.}
In order to explain our choice, let us return to 
\fref{fig2}. The two nuclear species there emphasized, 
 $^{12}$C and $^{14}$N, have small S-factors and 
 give rise to  the slowest reactions of the CNO-I cycle; however, 
the $^{14}$N reaction is much slower,  due to  Coulomb suppression.
When the cycle begins, N-neutrinos are soon produced and then $^{14}$N is synthesized.
This is all that happens
in the external part of the solar core, 
where the temperature is not too high: {\bf only half-cycle is active,} only N-neutrinos are produced--see e.g., Fig.~2b of\cite{vito}
and Fig.~1.2 right panel of.\cite{nuria}
In the  inner part of the solar core, instead, the CNO-I cycle proceeds unimpeded and the N- and O-neutrinos are equally produced.

\medskip
\noindent 
{\em Summary:} 
Eqs.~(\ref{kuly},\ref{kuly2},\ref{cciefa},\ref{cciefa2}) agree perfectly with 
\eref{cita1}; in other words, the procedure advocated by Bahcall, {\em with the new coefficients,} leads to the same result that follows from  
physical considerations.   
The result, \eref{cita1}, 
can be presented in compact form and in close  resemblance to 
 \eref{clitoi00},
\begin{equation} \label{citabo}
\color{blue}{ \sum_i \Phi_i \left( 1 - \frac{2 \langle E_i \rangle}{\mathcal{E}_{\mathrel{\mathsmaller{\mathsmaller{\odot}}}}}  \right)
- 0.123\times \delta\Phi_{ \mbox{\tiny CNO}}  = 
\frac{6.379\times 10^{10}}{\mbox{\small cm$^2$s}} (1\pm 0.4\%)}
\end{equation}
compare with Eqs.~(\ref{musu},\ref{papar}) and 
see \eref{clita}  for the $ \delta\Phi_{ \mbox{\tiny CNO}}$-term; 
note that according to the SSM, 
the $ \delta\Phi_{ \mbox{\tiny CNO}}$-contribution is below the present accuracy.
\Fref{fig2} and  \ref{fig3} remind us the 
physical conditions from nuclear physics,
\begin{equation}
 \Phi_{i}\ge 0\ , \ \Phi_{ \mbox{\tiny pp}}+\Phi_{ \mbox{\tiny pep}}\ge \Phi_{ \mbox{\tiny Be}}+\Phi_{ \mbox{\tiny B}}+\Phi_{ \mbox{\tiny hep}}\ , \ 
 \Phi_{ \mbox{\tiny N}}+ \Phi_{ \mbox{\tiny F}}\ge  \Phi_{ \mbox{\tiny O}}\ge  \Phi_{ \mbox{\tiny F}}
\end{equation}


\section{Flavor transformations of solar neutrinos}\label{sec2}
Before using the luminosity constraint, it is necessary to discuss a bit  
the interpretation of the observations of solar neutrino telescopes. 

The original goal  was to measure neutrino fluxes in order 
to understand the Sun, taking full advantage of  the SSM (i.e., using and/or testing it).
This goal is still valid and it is  still pursued. 
It inspired the concept of the first solar neutrino experiment, Homestake, and
of the most recent one, Borexino. 

However, the experiments proved that solar neutrinos are modified by peculiar phenomena.
Indeed, when neutrinos are produced in the Sun they have electronic flavor,  
but subsequently, when they reach the Earth,  they are a combination of the various neutrino flavors--as first argued by Pontecorvo.
An entire generation of solar neutrino observatories has worked to clarify the nature of these phenomena. 

Nowadays the laws of flavor transformation are known and verified with independent terrestrial experiments, in particular with KamLAND.
The most popular opinion is that these phenomena  can 
be precisely described within a reasonable particle physics model where the three known neutrinos are endowed with mass.
Thus, it is possible to return on the original goal, and use the solar neutrino experiments (along with the knowledge of 
flavor transformation phenomena) to reconstruct the emitted fluxes.\footnote{Note that, strictly speaking, only the SNO experiment has validated the predictions of the SSM  by measuring directly one flux--that of B-neutrinos. Their result does not depend upon the details of the three flavor transformations.
However, it is fair to remember that H.~Chen himself was inspired by the predictions of the SSM.\cite{chen}\label{fota}}

This is the position that we take in the rest of the present work:  
{\em flavor transformation phenomena are supposed to be 
understood.} 
In the appendix there is a somewhat broader discussion of the 
particle physics picture, with a brief mention of alternative models, which have some motivation but lack sufficient evidence to be considered too seriously for now. See also.\cite{vs1,vs2}

\section{Applications}\label{sec4}
In this section, we consider various applications of the luminosity constraint, focussing in particular 
on its uses for the purpose of extracting the CNO signal--presumedly, the hottest issue in solar neutrino physics.\cite{vs1}

\subsection{A quasi-empirical model for the PP-chain}
 \tref{tab2} resumes the {\em direct information} on the fluxes of  the PP-chain, where,\\
$\bullet$ the results on the pp-neutrinos obtained from the  
Gallium experiments\cite{sg09} are averaged 
with the direct determination  of Borexino;\cite{bx17}\\
$\bullet$ the values of the Be- and pep-neutrinos are as measured by Borexino;\cite{bx17}\\
$\bullet$ for the latter, the results from the two most recent SSM (with different metallicity) are averaged, 
enlarging the quoted error accordingly; \\
$\bullet$  the B-neutrinos flux measured by SNO is used (this is independent from 
assumptions on 3 flavor neutrino transformations, see again 
footnote \ref{fota});\\
$\bullet$  the hep-neutrino bound as derived by SNO\cite{sno} and Super-Kamiokande\cite{sk17} is cited
(SNO has a new analysis of hep-neutrinos in preparation);\\
$\bullet$  symmetric uncertainties  are given 
for the purpose of illustration, even if  
their values should be considered with a grain of salt.

The \textit{quasi-empirical model} of  \tref{tab2}  was defined as follows,\newline
(1) the CNO-neutrinos are set to zero by definition; \newline
(2) the precise value of pp-neutrinos follows from luminosity constraint,  
by applying a straightforward $\chi^2$ analysis;\newline
(3) the ratio pep/pp fluxes was set to the theoretical value of
$2.35\times 10^{-3}$ of,\cite{adelb11} including conservatively (but to some extent arbitrarily) a 2\% error;\newline
(4) the values of the Be- and pep-neutrinos stay unchanged; \newline
(5) the value of the hep-neutrinos is set to the theoretical value,
which  incidentally gives a good fit to  Super-Kamiokande data.\cite{sk17}\newline
This model is useful, since 
the luminosity constraint has a great impact on pp-neutrinos:
the errors on the 
pp and the pep-neutrino fluxes decrease 
by one order of magnitude.
The first hypothesis was considered also in the past\cite{spiro} with 
different motivations; to date, the merit it has is of breaking the entanglement, 
that will be discussed in the next section.

\begin{table}[t]
\tbl{Fluxes of the PP-chain and their uncertainties. 
$2^{\mbox{\tiny nd}}$ and $3^{\mbox{\tiny rd}}$ lines: current 
observational information.
$4^{\mbox{\tiny th}}$ and $5^{th}$ lines: quasi-empirical model. 
We adopt the same notation as in \eref{ntn} and in \tref{tab1}.
 \label{tab2}}
{\begin{tabular}{c|ccccc}
                                         ident.\ index $i$    & {pp} & ${\mbox{Be}}$ &  {pep} & ${\mbox{B}}$ & {hep}   \\ \hline\hline
                        observation  & 6.04 & 4.99 &  1.33 & 5.25 & $<23$   \\ 
      $\mbox{\tiny errors}$           & \color{blue}{\bf 0.53} & 0.13 &  \color{blue}{\bf 0.29} & 0.20 & 90\%  \\   \hline 
                                quasi-empirical & 6.02 & 4.99 &  1.41 & 5.25 & 8   \\ 
      $\mbox{\tiny errors}$           & \color{blue}{\bf 0.03} & 0.13 &  \color{blue}{\bf 0.03} & 0.20 & 9   \\   \hline  
\end{tabular}}
\end{table}

\subsection{Entanglement of PP and CNO-neutrino signals:} The 
luminosity constraint fixes precisely a linear combination of the various neutrino fluxes; the largest contribution in the SSM
is due to the pp-neutrinos, followed by the one of the Be-neutrinos. In order to illustrate this point better, and 
using the notation of \eref{ntn}, namely, $\Phi_i=\varphi_i\times 10^{\alpha_i}$/(cm$^2$s),  
we explicitly include the order-of-magnitude factors 
(i.e., the values of $\alpha_i$ from \tref{tab1}) obtaining 
\begin{equation}
\begin{array}{c}
0.9800 \times  \varphi_{\mbox{\tiny pp}} + 0.0939 \times \varphi_{\mbox{\tiny Be}} + 0.0092 \times   \varphi_{\mbox{\tiny CNO}} +\\
+ 0.0089 \times   \varphi_{\mbox{\tiny pep}} +\mbox{\small small terms}  = 6.379 \times (1\pm0.4\%)
\end{array}
\end{equation}
The coefficient of the CNO-neutrinos depends upon the SSM in principle, 
but in practice it does not vary much--in agreement with the discussion after \eref{citabo}.

This constraint can be exploited together with two recognized facts about solar neutrinos: 
1)~The Be-contribution is measured so 
precisely in Borexino,\cite{bx17}\ that can be considered known; 2)~the 
pep-contribution is strictly linked to the pp-one.\cite{adelb11} In this manner we find,
\begin{equation} \label{obo}
\color{blue}{  \Phi_{\mbox{\tiny pp}} + 0.93 \times   \Phi_{\mbox{\tiny CNO}}  = 6.02\times {10^{10}}{\mbox{(cm$^2$ s)}} \times (1\pm0.5\%)}
\end{equation}
that improves\cite{spiro,sidney}.
Thus, the central values of pp- and CNO-neutrino fluxes 
are \underline{almost fully anticorrelated,} since it is just the above combination which is determined by the luminosity constraint.
In fact,  if we had not measured the neutrinos (but only the light), it would impossible to conclude on a firm observational ground   
that the Sun works mostly by the PP-chain.

\subsection{Radiochemical experiments and the CNO-neutrinos}
The measured rate at the Homestake (chlorine) experiment is
\begin{equation}
R_{\mbox{\tiny exp}}(\mbox{Cl})=2.56\pm 0.23 \ \mbox{SNU}
\end{equation}
The contributions in the {\em quasi-empirical model}, in order of importance, are,
\begin{equation}
\{\mbox{B},\mbox{Be},\mbox{pep,hep}\}=
\{ 1.86, 0.64, 0.12, 0.01\}\ \mbox{SNU}
\end{equation}
The error on the cross section is estimated to be 
3.7\% for the high energy branches, B and hep, while it is 2\% for the other neutrino fluxes;
\begin{equation}R^{\mbox{\tiny PP}}_{\mbox{\tiny  th}}(\mbox{Cl})=2.63\pm 0.08\mbox{ SNU}\end{equation}
namely it amounts to 3.2\%.  
Therefore the rate of CNO-neutrinos, extracted from these data, is,
\begin{equation}R^{\mbox{\tiny CNO}}_{\mbox{\tiny  th}}(\mbox{Cl})=-0.07\pm 0.24\end{equation}

The measured rate from  
SAGE and Gallex/GNO is
\begin{equation}
R_{\mbox{\tiny exp}}(\mbox{Ga})=66.1\pm 3.1 \ \mbox{SNU}
\end{equation}
Expectations based on  {\em quasi-empirical model} are in this case,
\begin{equation}
\{\mbox{pp}, \mbox{Be}, \mbox{B}, \mbox{pep}\}=
\{39.1,18.1, 3.9, 1.4\}\  \mbox{SNU}
\end{equation}
The error on the cross section obtained using\cite{sg09} (but see also\cite{sg17}) implies,
\begin{equation}
R^{\mbox{\tiny PP}}_{\mbox{\tiny  th}}(\mbox{Ga})=62.6^{+3.7}_{-1.9} \ \mbox{SNU}
\end{equation}
absolute errors of pp, Be, B are almost equal and are summed linearly--not quadratically. 
Assuming, 
$R_{\mbox{\tiny  exp}}(\mbox{Ga})= 
R^{\mbox{\tiny PP}}_{\mbox{\tiny  th}}(\mbox{Ga})+R^{\mbox{\tiny CNO}}_{\mbox{\tiny th}}(\mbox{Ga})$
we find
\begin{equation}
R^{\mbox{\tiny CNO}}_{\mbox{\tiny th}}(\mbox{Ga})=
3.5^{+3.6}_{-4.8} \ \mbox{SNU}
\end{equation}
To improve, more data are needed\footnote{E.g., the new 
$\sim 10$ yr of SAGE, presented at this conference.\cite{dre}} and much more importantly, the 
theoretical error should be lessened by {\em measuring the cross section.}

\subsection{Summary of what we know on CNO-neutrinos}
The expected CNO-neutrino flux, deduced  from \tref{tab1} and \eref{mab}, is,
\begin{equation}
\varphi_{\mbox{\tiny CNO}}=4.88 \mbox{ [GS98]}, 3.51 \mbox{ [AGSS09]}
\end{equation}
It depends upon the SSM.
Borexino has searched for CNO-neutrinos:
The 95\% bound 
$\varphi_{\mbox{\tiny CNO}}<7.9$ 
was obtained 
scaling together the three fluxes, as predicted by the most recent SSM.\cite{bx17}
In their fig.~6 one finds the $\Delta\chi^2_{\mbox{\tiny Borexino}}(n)$ as a function of the counting rate $n_{\mbox{\tiny CNO}}$, obtained assuming the GS98 model. This is well-described by a parabolic shape, which implies that the likelihood is almost Gaussian. 
The minimum is at $n_{\mbox{\tiny CNO}}=2.4$ counts per day/100t, that corresponds to a flux $\varphi_{\mbox{\tiny CNO}}=2.34.$\cite{bx17} Therefore, 
we have,  
\begin{equation}
\Delta\chi^2_{\mbox{\tiny Borexino}} (\varphi_{\mbox{\tiny CNO}})=\frac{(\varphi_{\mbox{\tiny CNO}}-2.34)^2}{\delta \varphi^2}
\mbox{ with } \delta\varphi=2.85\end{equation} 
where the value of $\delta\varphi$ reproduces 
the quoted 95\% bound and where we assume that the best fit value does not change drastically assuming the AGSS09 model instead.
Summarizing,  Borexino allows us 
to derive important empirical conclusions on the CNO-flux: 
\begin{enumerate} {\small
\item its value is {\bf lower} than the one indicated by the GS98 and AGSS09 models;
\item within the upper 1$\sigma$ range, it is compatible with both of them; 
\item within the lower 1$\sigma$ range, it is compatible with no CNO flux at all.

}
\end{enumerate}
By combining the results of gallium experiments and Borexino, we find (for both SSM) a very mild shift upward 
$
\varphi_{\mbox{\tiny CNO}}=2.69\pm 2.77
$, 
that we quote as,
\begin{equation} \label{cagatur}
\color{blue}{ \Phi_{\mbox{\tiny CNO}}= ( 2.69\pm 2.77 ) \times \frac{10^8}{\mbox{\small cm$^2$s}}}
\end{equation}
The important  conclusions (1), (2), (3), outlined just above, remain valid.
All this shows that the inclusion of the results of the gallium experiments in the analysis 
does not modify strongly {\em the Borexino limit on CNO-neutrinos, which is the main information available today from the observations.} 

\subsection{Future chances in Borexino and pep-neutrinos}
As emphasized by D.~Guffanti,\cite{dre}, in 
the region of energies where CNO-neutrinos can give an observable signal,  pep-neutrinos are present: Thus, this ``beam-related background'' should be known as precisely as possible, in order not to interfere with the extraction of the CNO bound (or signal).

One procedure that does not imply the theoretical input of the SSM, but only the usage of the luminosity constraint, is simply the following one:
We can perform a global fit to the data using the luminosity constraint to determine the pp-neutrino flux accurately, moreover we can employ the known ratio between pep and pp\cite{adelb11} to fix pep-neutrinos reliably.

The theoretical error on this ratio is presumedly small; still, it would be useful to assess 
precisely its value in future.
Indeed, even if the matrix element of the two reactions leading to pp- and pep-neutrinos is the same, $(i)$~pp- and pep-neutrinos are produced in slightly different regions, and $(ii)$~the description of electron capture requires theoretical modeling. 

\section{Discussion}\label{sec5}
Great results in 
solar neutrino astronomy have been obtained and new ones are expected:
There is also a chance of measuring for the first time 
a signal from CNO-neutrinos, after those seen from the PP-neutrinos, 
thereby elucidating the two main astrophysical mechanisms that fuel stars.

The measured solar luminosity, with minimal theoretical inputs, leads to the {luminosity constraint}, 
that is based on the assumptions that  the Sun is in equilibrium and 
we understand sufficiently well nuclear physics. 
This is a precious tool to proceed further in the study of the Sun; we have  proposed an improved description
and discussed it thoroughly.

The luminosity constraint and several other facts  imply 
that the {\em PP and CNO-neutrino signals  are entangled}, due to the empirical need to extract both of them from 
the experimental data. The apparently simplicity of this point should not lead us to underrate its importance; instead, it 
should be taken into account attentively in order to plan future steps forward at best.

We have examined various consequences of this situation:  
1)~The entanglement implies that it is momentous to probe CNO-neutrinos; Borexino has the best chances to fulfil this goal and it will indicate us the way to go. 
2)~The knowledge of the gallium cross section should be improved; its uncertainty hinders  the chances of exploiting at best existing measurements.
3)~Further work to measure the solar luminosity and to assess the connection between pp- and pep-fluxes would be desirable.

\section{Acknowledgments}
I thank M.~Meyer and K.~Zuber for the invitation, the  {\em German Alumni} programme of TUM Dresden for support and in particular M.~Richter-Babekoff and her staff for the very professional and friendly assistance.
I am grateful to C.~Mascaretti for collaboration in the early stage of this work and  also to G.~Bellini, M.~Busso, I.~Drachnev, M.~Junker, A.~Gallo Rosso, T.~Kirsten, V.~Gavrin, D.~Guffanti, S.~Marcocci, L.~Marcucci, L.~Pandola, G.~Ranucci, A.Yu.~Smirnov, D.~Vescovi and D.~Xue-Feng for pleasant discussions.

\appendix{On flavor transformation of solar neutrinos\label{aref}}

The phenomena of neutrino transformations are often called 
`neutrino oscillations' see e.g.\cite{msw,msw2} even when they do not have oscillatory character.
This is true for solar neutrinos,\cite{alex} thus we refrain to use this terminology.
Here, the reference model for these phenomena, based on the assumption 
that the three known neutrinos have small masses, is summarized, discussing briefly alternative schemes of interpretation.
See the talks at {\sc Neutrino 2018}\cite{niu} for updated information and also\cite{vs1,vs2} 
for a discussion that emphasizes the connections of neutrino transformations and  solar neutrino astronomy.

\noindent
{\bf The three flavor interpretation:}
It is a recognized fact that under suitable conditions neutrinos  are subject to flavor transformations. The current interpretative framework can be summarized as follows:
1)~there are 3 light species of neutrinos subject to weak interactions, $\nu_{\mbox{\tiny e}},\nu_\mu,\nu_\tau$;
2)~consistently with solar and atmospheric neutrino data, they undergo flavor transformations, i.e., 
the final flavor is not the initial one;
3)~these phenomena are attributed to neutrino masses, 
and the parameters  relevant for their description 
have been measured;\cite{lisi}
4)~this model 
was tested by means of reactor and accelerator 
experiments;  
5)~the assumption on the number and on the mass of the neutrinos is consistent with observational cosmology, namely, what we know
from  the $\Lambda$CDM cosmological model on 
big-bang nucleosynthesis and cosmic microwave background.

The current theory of solar neutrino transformations has been established since more than 30 years, when the matter - or  MSW\cite{msw,msw2} - effect was finally understood. 
Let us recall the main features of this theory: The hamiltonian of neutrino propagation includes, besides 
the term due to neutrino masses, also a further term--only for electron neutrinos--due to weak interactions with the electron in the medium: This is 
the matter term. 
Due to the matter term, the high energy solar neutrinos--such as the B-neutrinos--are produced as local-mass-eigenstates in the core of the Sun and exit as such, leading to the survival probability $P_{ee}=|U_{e2}^2|\sim 0.3$. 
The low energy  solar neutrinos - as the pp-, Be- and pep-neutrinos - do not feel strongly the effect of the matter term
and this leads to (phase averaged) flavor transformation in vacuum, $P_{ee}=\sum_{i=1}^3 |U_{ei}^4|\sim 0.6$. 
This was reviewed in details by A.~Smirnov,\cite{dre}
discussing the formalism, the characteristic features and other manifestations, 
either observed or observable.

The parameters of flavor transformation indicated by 
a global interpretation of the data\cite{lisi}--and more specifically the $\Delta m^2_{12}$ measured by the KamLAND experiment--lead to two expectations: a)~the spectral shape should be distorted, so to grow at low energies--`upturn';
b)~the solar neutrinos that pass through the Earth, that have already undergone flavor transformation in the Sun,  should undergo a small conversion back into electron neutrinos--`regeneration': but `upturn' is not seen while `regeneration' is larger than expected, see \cite{sk17} and compare with Y.~Suzuki.\cite{dre} 
Thus,  even if 
the observed phenomena do not have a strong significance to date,
the very detailed analysis of the B-neutrino flux of Super-Kamiokande suggests caution.

\begin{figure}[t]
\begin{minipage}[c]{5.6cm}
\caption{The oscillation probabilities obtained by the rates of SNO and Borexino 
lead to a best fit point 
- white circle dotted in black - that agrees very well with the global fit point - white circle.
The point indicated by Super-Kamiokande measurements - with `upturn' and `regeneration' - 
is also shown for comparison - black circle dotted in white.
From.\cite{vs2}
}
\end{minipage}
\hskip5mm
\begin{minipage}[c]{4.3cm}
\includegraphics[width=2.0in]{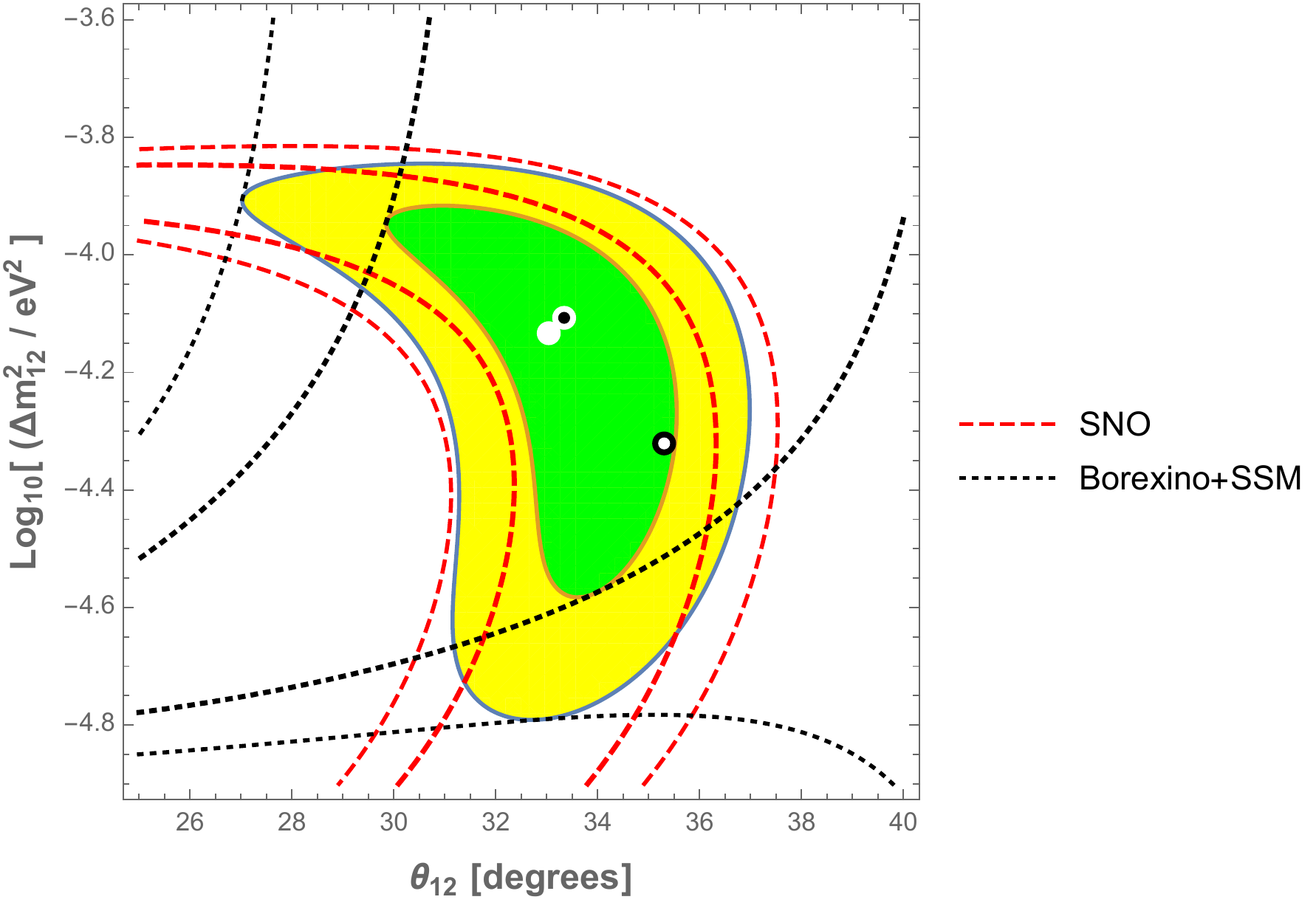}
\end{minipage}
\label{figQuat}
\end{figure}

On the other hand, it has been noted that, SNO and Borexino solar neutrino data allow us to reconstruct the `survival probability' of the MSW theory, even omitting Super-Kamiokande data altogether; proceeding in this manner one finds a best fit point  that agrees perfectly with the one indicated by the global analyses and by KamLAND\cite{vs2} as shown in \fref{figQuat}.

\noindent
{\bf More on the r\^ole of the theory:}
In view of the historical discussions at our meeting,\cite{dre} it is appropriate to conclude with a few considerations on 
the impact of theoretical considerations in the discussion of solar neutrinos. 

The `standard model' of elementary particles led us to consider with favor the idea of three families, which implies the existence of three massless neutrinos subject to weak interactions. 
The most direct extension of this model includes non-renormalizable operators suppressed by large mass scales, that describe the effects of new physics\cite{wei} and endows the neutrinos of the standard model with small but non-zero masses. 

The ensuing hypothesis of three light and massive neutrinos has been discussed and compared with the available data since the beginning. It 
results into a {\em predictive} model for flavor transformations: E.g., in 2001 there were ambiguities in the allowed 
parameters:  using this model it was argued\cite{2001} that KamLAND and Borexino were enough 
to resolve them by 
measuring the relevant parameters, as it happened eventually.

The existence of {\em other} neutral fermions without weak interaction (e.g., `right-handed neutrinos) seems to be 
plausible; the real question concerns however the value of their mass, a parameter that in the context of the standard model has nothing to do with the electroweak scale and thus it is completely undetermined.
A popular and reasonable position is that their mass is fixed by new gauge interactions at much higher scale. 
The converse assumption that some of these fermions is very light (and could play a r\^ole in the observed phenomena of flavor transformation)  
cannot be fully excluded, even if it lacks of convincing 
theoretical motivation to date. 

Let us consider however the existence of a very light fourth neutrino, without weak interactions: 
a `sterile' neutrino.
Assuming that such a hypothetical particle has a non-negligible mixing with the other neutrinos,
various individual anomalies  can be addressed.  However, no global  analysis 
that has adopted a precisely defined model of this type has found significant evidence.  
Already the first one failed to find any significant hint.\cite{cir}
A recent global analysis of M.~Maltoni\cite{niu}  indicates internal contradictions.  
This shows, once more, that {\em well-formulated 
models are useful} to interpret the data.
Moreover, direct tests of the so-called `gallium anomaly' and `reactor anomaly', within simplified 2 flavor schemes, lead to disagreement with the DANSS, Stereo and NEOS experiments.\cite{niu}  

Finally, one could speculate about new interactions felt only by neutrinos, or maybe by the new hypothetical neutrinos,  that can lead to further matter effect, aka, non-standard interactions (NSI),  as first argued  in.\cite{roulet,serg}
This can help to address current anomalies of some experiments, however it is not clear that  
well-defined minimal models are viable, when one considers that the new interactions should show up in the phenomena of interest and  also elsewhere, say, in collider or flavor experiments. In fact, the known neutrinos are part of leptonic doublets, e.g., $\ell_e=(\nu_{\mbox{\tiny e}},e)$, and the assumption of NSI implies also phenomena involving charged leptons.  

Let us summarize: 
We have assumed 
the conventional 
3 flavor description of neutrino transformation phenomena
all throughout the present work.
This is a description that has provided and that can provide us valid guidance in the discussion of the experimental findings. Different opinions follow, when excessive simplifications are adopted for the description of the data, or conversely when complicated theoretical schemes are adopted despite the weakness of their current motivations.  For more discussion, see.\cite{vs1,vs2}

{
}

\end{document}